# The 2020s political economy of machine translation

Steven Weber

School of Information and Department of Political Science

University of California, Berkeley    USA

102 South Hall  UC Berkeley Berkeley CA USA 94720

M: +1 415.203.8432     O: +1 510.643.3755

[steve_weber@berkeley.edu](steve_weber@berkeley.edu)

Steven Weber is Associate Dean and Professor at the School of Information and Professor in the Department of Political Science at UC Berkeley. He directs the Center for Long Term Cybersecurity at Berkeley. His most recent book is *Bloc by Bloc: How to Organize a Global Enterprise for the New Regional Order* (Harvard University Press, 2019).

**The 2020s political economy of machine translation**

This paper explores the hypothesis that the diversity of human languages, right now a barrier to 'interoperability' in communication and trade, will become significantly less of a barrier as machine translation technologies are deployed over the next several years. I argue that machine translation will become the 2020's analogy for ideas to what container shipping did for goods trade in the second half of the 20th century. But as with container shipping or railroads in the 19th century, this new boundary-breaking technology does not reduce all boundaries equally, and it creates new challenges for the distribution of ideas and thus for innovation and economic growth. How we develop, license, commercialize, and deploy machine translation will be a critical determinant of its impact on trade, political coalitions, diversity of thought and culture, and the distribution of wealth.

Keywords: trade, globalization, machine translation, inequality, productivity

This work was supported in part by the Hewlett Foundation under a grant to the Center for Long Term Cybersecurity at Berkeley; and by the Berggruen Institute (Los Angeles) where the author is a Berggruen Fellow.

The most important and most desirable technology revolutions are those that boost economic productivity and improve the human experience at the same time. That normative assertion may sound obvious, but historically it sets a high bar. The 18th century Industrial Revolution would not qualify because for most people outside of factory-owning capitalists, productivity rose at the cost of a largely degraded human experience in many aspects of life. The introduction of electricity probably comes closer to qualifying, as does the printing press. There's long been doubt about whether digital technologies, computing and the Internet meet these criteria. The productivity gains are unmistakeable, but the overall impact on human experience is now generally seen as ambiguous and possibly net negative.

Of course technology by itself is neither inherently good nor bad, but rather how we understand, organize, deploy, pay for and incorporate technology is what determines its impact. Those elements are largely under the control of people, but are much easier to maneuver at early stages rather than later in the life-cycle of any particular technology.

The core proposition of this paper is that machine translation is just now entering the early 'sweet spot' period where thoughtful decisions can tilt the table toward better impact for both economic productivity and human experience. A subsidiary proposition is that because machine translation runs on a global computing infrastructure, it is inherently necessary to think from the start about its global impact (the steam engine took decades to diffuse from its British home to the rest of the world).

The broadest ambition should be to develop and deploy machine translation technology in a manner that is Pareto-improving across five dimensions of liberal progress: enhancing productive competition, ensuring fundamental fairness, guaranteeing a baseline of protection for the most vulnerable, deepening pluralism and stimulating innovation.[1] A narrower goal is to aim simply for greater equality in the distribution of economic gains, as a necessary but not sufficient condition.

Put differently, the argument is machine translation will not contribute to advancing global liberal outcomes if it reinforces (or supercharges) contemporary trends toward imbalanced distribution of economic gains from digital technology. But in the absence of intentional intervention, I will argue that this is exactly what is set to happen. We can do better, and the normative goal of this research is to reason toward specific proposals that would improve both productivity and distributional equity at the same time.

**Language Matters**

Human language is central in the coordination of complex economic activities that generate productivity. Consider a (possibly fictitious) example from the Biblical story of the Tower of Babel:

---

[1] These five dimensions are drawn from forthcoming work with Naazneen Barma and Brent Durbin, *How to Make the World More* Liberal

> 'now the whole world had one language and a common speech… they said, come let us build ourselves a city with a tower that reaches to the heavens that we may make a name for ourselves'.

Long before economics had concepts of transaction and coordination costs, the biblical story had a deep intuition about the importance of language, written and spoken, in both. The bible over-estimated what bricks and mortar could actually do, but it does recognize the potent combination of human ambition and the ability to coordinate. The story might be seen as an early version of the modern dictum 'if we really work together there is nothing we cannot do'. In this case, having a single language would enable coordinating the labour of all humankind so that people could build a tower to heaven and come close to being on the same level as God.

God didn't much like that idea. And so:

> 'the Lord came down to see the city and the tower the people were building. The Lord said, if as one people speaking the same language they have begun to do this, then nothing they plan to do will be impossible for them. Come, let us go down and confuse their language so they will not understand each other'

Thus the introduction of language-based transaction costs friction into economic coordination, which 'scattered the people over the face of the whole earth'. If groups of people couldn't communicate with others to coordinate production, there wasn't much reason for them to stay in the same physical location. People who spoke the same language went off in groups to inhabit different parts of the earth; production of towers and other things reverted to smaller scale coordination challenges; and distinctive cultures developed in somewhat insulated environments where ideas from 'inside' a particular language could diffuse orders of magnitude more quickly and easily than ideas from an 'outside' language silo that was both physically and linguistically separate.[2]

The Babel story highlights a few important arguments — upside and downside — about the multiplicity of human languages. First, even moderately complex economic coordination requires granular communication. Introducing language friction into that process is a significant transaction cost in itself, and probably magnifies the impact of other transaction costs at the same time (for example, costs associated with imperfect contracting likely are made worse when contracts are rendered in multiple languages). Second, language barriers have a protective effect on the evolution of different ideas, practices, religions and culture, that might otherwise be subject to homogenization or at least consolidation if those barriers disappeared (more on that later).

Third, the impact on overall productivity of a multiplicity of languages is at the highest level a consequence of how these two vectors intersect. More diversity of ideas should contribute positively to innovative potential and so would help increase productivity (multiple experiments going on in parallel accelerating useful discovery); but the language barrier to coordination be-comes a drag on learning from experiments in other places where what is discovered gets encoded in a different language. It also constrains the ability to scale.

---

[2] The analogy here is to 'evolutionary isolation' as explained by Charles Darwin's study of the Galapagos Islands

There's no a priori argument to determine which vector dominates over time, but intuition along with a casual reading of history suggests that learning across cultures and languages is much slower than it 'ought' to be; that mistakes are repeated and dead-ends pursued much more often than they should be; and that increased diversity of ideas and practices relevant to economic productivity isn't sufficient to overcome those negatives. If that intuition is roughly accurate, then overall economic productivity grew more slowly because of language diversity (which is, after all, what God intended in the story). That's not a comment on whether cultural and other forms of diversity are 'worth it', it's simply a hypothesis that productivity would have been higher in an alternative history where – all other things equal – human beings spoke the same language.

All other things equal is a massive qualifier, that science fiction stories have played with for decades. Gene Roddenberry in the original 1960s Star Trek series equipped the crew of the Starship Enterprise with a hand-held cylindrical universal translator that Captain Kirk used to communicate with newly discovered species. It was partly because the technology worked as well as it did that the Federation established Starfleet General Order 1 (this order, also known as the Prime Directive, prohibited Starfleet explorers from interfering with the internal and 'natural' development of alien civilizations).³ It factored into the series particularly when the Enterprise encountered aliens that appeared to be in a more primitive state of development than the Federation, the presumption being that less technologically sophisticated civilizations would be subject to cultural-ideological domination (even unintentionally), *when their language could be fully understood* (in both directions, hearing and speaking) by the crew of a Starship.

Apart from that limitation, the Star Trek universal translator played an important role in resolving misunderstandings and avoiding what could have been violent conflicts. In stark contrast, *The Hitchhiker's Guide to the Galaxy* (Adams, 1979) proposed that universal translation could be the most important cause of future wars. This was because the books' 'Babel fish', which inserts into the user's ear and translates by 'reading' the mental frequencies of the speaker, allows clear and transparent understandings of the actual level of hostility and conflict between people and societies, which before presumably had been hidden to some extent behind the barriers of incomprehensible speech.⁴

---

³ In further works that elaborated on the show, this articulation of the Prime Directive became canon: As the right of each sentient species to live in accordance with its normal cultural evolution is considered sacred, no Starfleet personnel may interfere with the normal and healthy development of alien life and culture. Such interference includes introducing superior knowledge, strength, or technology to a world whose society is incapable of handling such advantages wisely. Starfleet personnel may not violate this Prime Directive, even to save their lives and/or their ship, unless they are acting to right an earlier violation or an accidental contamination of said culture. This directive takes precedence over any and all other considerations, and carries with it the highest moral obligation. See Menke, B. E. & Stuart, R. D. (1986). *The Federation*. Chicago, IL: FASA. pp. 5.

⁴ per Adams (1979), 'The poor Babel fish, by effectively removing all barriers to communication between different cultures and races, has caused more and bloodier wars than anything else in the history of creation' (pp. 61).

These two hypotheses about the conflict-promoting or conflict-avoiding effects of translation map almost exactly onto long-standing arguments in international relations theory about interdependence between countries. Commercial liberalism posits that higher levels of trade between countries bind them together in valuable relationships that conflict would place at risk, and thus interdependence makes conflict less likely. Structural realism, in contrast, argues that high levels of interdependence are actually a source of conflict (over distribution of gains, technology appropriation and so on) and that lower levels of interdependence are more likely to be associated with peace because countries that are less dependent on each other, have less opportunity to try to use the resulting leverage against each other and have less to fight about overall.[5]

Rather than one theory being right and the other wrong, what is really happening is that one or the other causal mechanism predominates under different sets of conditions. It's reasonable to posit that the same is likely to be true of translation technologies — that the overall effect of translation on conflict (as one example of an outcome) will be some kind of sum of multiple causal mechanisms and thus depend on conditions that amplify or weaken upside and downside vectors.

That's a hopeful inference, because some of those conditions are going to be subject to decisions that people make about how the technology is developed, deployed, licensed and paid for etc. But before we can get more specific about what those conditions and decisions actually look like, we need to examine some of the constraining affordances that modern machine translation technology operates under. How the technology actually works and what those mechanisms do and do not make possible in the medium term future set the stage for what decisions can be realistically framed up.

---

[5] See for example Michael W. Doyle, "Liberalism and Foreign Policy," Chapter 3 in *Foreign Policy:Theories, Actors, Cases,* Steve Smith, Amelia Hadfield, and Timothy Dunne eds, Oxford University Press 2016. The microfoundations of these effects have been hypothesized in more elaborate ways over time (such as multinational firms who depend on global supply chains lobbying governments against paths to conflict).

**'Universal' isn't Universal**

Ethnologue currently estimates that there are over 7000 known living languages, many of which are used by very small numbers of people In comparison, Google's Cloud Natural Language API recognizes less than 100 languages; and as of February 2020 offers syntactic analysis on 11 and sentiment analysis on 10.[6]

      The perennial goal of universal machine translation is to create a single model that translates between any and all language dyads with equally high accuracy. Today's technology isn't close. A casual user can see some of the limitations by playing with Google Translate on any web browser. Google Translate offers about 100 languages; it works extremely well translating between English and other common languages; moderately well between English and less commonly used languages; and less well between a pair of infrequently used languages. Some of the spoken language translation services (now available on mobile phones for example) are even

---

[6] Language Support [User Documentation]. Retrieved from https://cloud.google.com/natural-language/docs/languages

more limited. These systems seem almost magical if you are an English-only speaker working with French in Paris or Spanish in Madrid; they are much less useful if you are a Burmese-only speaker trying to communicate in Basque.

The simplest way to understand these limitations is to think of today's machine translation algorithms as supervised learning systems that are built on reliable and accurate labelled training datasets. As is often true, high-quality training datasets are out in the world as a coincidence of other needs and processes. Commonly used data sets for machine translation include TED talks (manually translated in up to 59 languages); documents from the European Parliament (which by law are translated into 21 European languages); and the UNCorpus (more than 11 million sentences rendered in 6 languages). The Bible Corpus is generally considered the broadest public available multilingual dataset (about 30,000 sentences in 900 languages).[7]

Linguists and machine translation scientists make an important distinction between what they call high-resource and low-resource languages. High resource languages are, expectedly, languages for which many data resources exist — English is by the far the highest resource lan-guage, with many Western European languages as well as Japanese and Chinese also reasonably high resource. Low-resource languages — including many languages from poorer countries, many local dialects, rarely-used or semi-extinct languages — make up the vast majority of the rest. There are a large number of languages that are mostly spoken and for which very few written resources exist. There are languages with a large amount of digital raw text from various genres (everything from social media to scientific papers) and lexical, syntactic and semantic resources (dictionaries, semantic databases) as well as various bodies of highly annotated text (for example, text that is labelled with parts-of-speech tags); and many languages that have little to no such resources available.

---

[7] Cettolo, M., Girardi, C., & Federico, M. (2012, May). Wit3: Web inventory of transcribed and translated talks. *EAMT 2012. Proceedings of the 16th EAMT Conference* (pp. 261–268). Trento, Italy; Koehn, P. (2005, September). Europarl: a parallel corpus for statistical machine translation. *MT Summit X. Proceedings of the 10th Machine Translation Summit* (pp. 79–86). Phuket, Thailand: AAMT; Ziemski, M., Junczys-Dowmunt, M., & Pouliquen, B. (2016, May). The United Nations parallel corpus v1.0. *LREC 2016. Proceedings of the 10th International Conference on Language Resources and Evaluation* (pp. 3530-3534). Paris, France: European Language Resources Association; Tiedemann, J. (2018, March). Emerging language spaces learned from massively multilingual corpora. *DHN 2018. Proceedings of the 3rd Digital Humanities in the Nordic Countries* (pp. 188-197). Helsinki, Finland: CEUR Workshop Proceedings.

This resource gradient among languages and the resulting disparity in machine translation capabilities is the most important technology variable shaping how machine translation will transform economies and culture. State-of-the-art neural machine translation (NMT) systems aspire to develop a single multilingual model that could handle all translation directions at once, rather than relying on a large number of dyadic models that could on aggregate, at least in theory, do the same (the number of necessary dyads would scale as a square of the number of languages; that many models would be almost impossible to train, deploy and maintain). Translating in sequential dyads (for example, going from Swahili to English and then from English to Spanish, in order to translate from Swahili to Spanish) is computationally intensive and less accurate, as some level of error is introduced at each sequential step.

A 2019 paper from Google AI surveys the state of universal machine translation.[8] What would make the problem easier would be if the learning signal from any particular language had a positive impact on the ability of the model to work with other languages — put differently, if the model were to generalize more effectively with each language that it learns. This is called positive transfer and it does appear to some degree in low resource languages; but the gains start to reverse themselves and performance on high resource languages start to decline after a certain point (this is called negative transfer or interference).

The massively ambitious model described in that paper uses an original dataset extracting parallel sentences from the web and contains 25 billion sentence pairs; and dataset training example sizes that range from around 2 billion for high resource language pairs down to 35 thousand for low resource language pairs, a difference of almost 5 orders of magnitude. The researchers experiment with a number of methods aimed at enhancing positive transfer and reducing interference (for example, over-sampling low resource language dyads to compensate for the imbalance in the training data). But there is no free lunch — oversampling on low resource languages improves transfer but creates interference that reduces performance significantly on high resource languages; regular sampling yields better retention of performance on high resource languages but sacrifices considerable performance on low resource languages. More sophisticated sampling strategies change the terms of these tradeoffs but do not eliminate them. The same is true for increases in model capacity that rely on better and more hardware and infrastructure.

---

[8] Arivazhagan, N., Bapna, A., Firat, O., Lepikhin, D., Johnson, M., Krikun, M, … Wu, Y. (2019) Massively Multilingual Neural Machine Translation in the Wild: Findings and Challenges. *arXiv*. Retrieved from https://arxiv.org/pdf/1907.05019.pdf.

What this adds up to is a technology weighted toward high resource language dyads and against low resource language dyads, with high-low dyads achieving translation capacity somewhere in the middle. Barring an unspecified technology breakthrough that would change the basic methods and remake the terms of the tradeoffs, it's reasonable to extrapolate substantial improvements in all translation tasks over the next decade, *but not all equally*. This imbalance - possibly increasing in relative magnitude — between translation among high resourced languages as compared to other dyads, will become a crucial feature of how this technology reshapes communication, commerce and culture on the global landscape.

**Historical Analogies and Interoperability**

Translation today is more like interoperability than it is integration or uniformity of language. Integration implies a frictionless state; uniformity implies a single language (like Esperanto). The IEEE defines interoperability as 'the ability of two or more systems or components to *exchange* information and to *use* the information that has been exchanged'.[9] The key distinctions are nicely clear when it comes to computer languages. Technical interoperability is the ability to open a file created in one application within a second application. Syntactic interoperability is a common format for data exchange (XML is an example). Semantic interoperability is the ability to under-stand the meaning of content in the same way by sender and receiver. Pragmatic interoperability is the ability to do things (coordination, for example) that follows from semantic interoperability and would not be possible without it.

---

[9] Interoperability. (1990). In *IEEE Standard Computer Dictionary: A Compilation of IEEE Standard Computer Glossaries*. IEEE.

These distinctions matter because they characterize the kinds of frictions that machine translation's partial interoperability capacity will reduce, leave in place and exacerbate or create anew. Not all aspects of interoperability depend on each other nor do they reinforce each other in linear ways. A casual but illuminating example that many people have experienced comes from the ability to navigate a subway map in a foreign country. In a recent trip to Japan with my partner who speaks a small amount of Japanese (I speak none), I was able to plan our subway trip around Tokyo much more easily than she. That is because I navigated the map (which was labelled only in Japanese) through syntax, interpreting the 'common form' of a subway map without translating a single label. She had the capacity for partial semantic interoperability, which led her to focus on the labels and actually *reduced* her pragmatic interoperability (mine was sufficient to get where we needed to go, without any semantic interoperability at all).

The same strategy might not work with pure text, but since machine translation will be applied to all kinds of artifacts that have words on them as well as pictures and diagrams (think of a technical manual for a machine tool) the story is still instructive. At the limit, the various kinds of interoperability might be synergistic with each other; but short of the limit (and where we will be in the next decade), it often won't be so.

This argument contextualizes the relevance of three simple historical analogies. The use of analogies here is a heuristic for starting to characterize challenges that machine translation will pose to global political economy. The first analogy is to what in the 1990s was called 'globalization' theory. In simple terms, the proposition is that language may now be as or more significant than physical distance as a barrier to interoperability in communication and trade across borders (i'll refine and evaluate that heuristic later).

The second analogy is about previous generations of technology that reduced the impact of those physical barriers in selective ways — notably railroads in the 19th century and container shipping in the 20th century. Both had hub-and-spoke topologies that shaped their impact on trade and as a consequence economic growth. This phenomenon is well documented in 19th century railroad routes and pricing schemes. It was, for example, very cheap to ship goods on rail from Chicago to St. Louis; more expensive to ship from St. Louis to non-hub small cities (like Dubuque Iowa); and very expensive to ship between two non-hub small cities (from Dubuque to Rock Island Illinois). This was partly determined by the technology of rail and partly by the business model of railroads, but the consequences were equivalent (at least until anti-trust authorities forced reform of what were determined to be discriminatory pricing schemes).[10]

Railroads vastly reduced the impact of physical distance *between particular nodes* in a network and, in relative terms at least, *increased* the barriers between others. This had the conse-quence among other things of further concentrating economic activity and exchange in the large city nodes and hollowing out of small towns and communities along the way.

---

[10] Gerald Berk (1997), *Alternative Tracks: The Constitution of American Industrial Order 1865-1917*, Baltimore MD: Johns Hopkins University Press.

The third analogy is about *The Mythical Man-Month* (Brooks, 1975) and what later became known as Brooks' Law (after the book's author Frederick W. Brooks).[11] Brooks was an observant software engineer at IBM in the 1970s who sought to explain why it seemed so difficult for teams to coordinate and cooperate effectively in the creation of complex software products. His view was distilled into Brooks' Law: that adding additional person-power to a late software project only makes it later. Brooks posited several mechanisms to explain this observation. The most obvious was the ramp up time for a new engineer to understand the work done before she joined the team (and the effort that team members would need to expend to help her get there).

More relevant to the machine translation challenge was Brooks' argument that communication overhead (or transaction costs, if you prefer that language) increases at a very fast rate when the number of people working together on a complex task rises. As Brooks put it, as more people are added to a team each person has to spend more time figuring out what everyone else is doing. That dynamic is less problematic for tasks that are easily divisible into discrete parts (which is of course Adam Smith's argument for a classical division of labour). But when tasks are highly interdependent or need to be precisely sequenced or really can't be divided up into discrete elements and then re-assembled, Brooks' Law says something profound about the va-garies of human communication and why the division of labour is not always a boost for productivity. As Brooks put it so cleverly, the fact that one woman can produce a baby in nine months does not mean that nine women working together can produce a baby in one month.[12]

---

[11] Brooks, F. P. (1975). *The mythical man-month: Essays on software engineering.* Reading, MA: Addison-Wesley.

Software engineering has progressed dramatically in both technical and sociological terms since Brooks wrote. Agile development processes and code repositories and Slack and lots of other innovations have softened the terms of Brooks' Law, but not eliminated the underlying insight about how hard it is for humans to communicate in words about abstract concepts and artifacts like software that exist almost on another plane, and at a level of complexity that no single person can visualize in their mind at once.

Machine translation will enter the picture with globalization dynamics, economic geography and Brooks Law still in place; the next section of the paper explores how these analogies and insights illuminate key challenges for political economy on the global stage.

**Challenges to Global Political Economy**

---

[12] The same dynamic operates in the opposite direction: The mythology says that Helen of Troy had such beauty that she could launch a thousand ships, but a milli-Helen would almost certainly not be the exact amount of beauty needed to launch one ship.

Machine translation at present poses at least three specific challenges to liberal progress in global political economy. The three I consider here are comprehension and false positives; diversity of thought and the power law; and technical barriers versus cultural barriers including elite differentiation within imagined communities.

*Comprehension and False Positives*

Shared language isn't equivalent to shared understanding, just as technical interoperability might not be semantic or pragmatic interoperability. Intuition and experience tell us that it is remarkably easy to misinterpret the meaning behind another person's speech even when they speak the same language. Most people have had decades to calibrate themselves in this respect and we rely on social signals and non-verbal cues to 'test' our understandings of what others have said in conversation. Even when we have a shared foundation of language, idioms and compatible cultural references, it's not easy and it's far from perfect — but we get it done with reasonable efficacy.

The first decade(s) of widespread machine translation will confront an environment where fewer or (sometimes) none of these shared foundations are present. The most profound human challenge of machine translation might very well be that people using the technology will *think* they understand more than they actually do about each other and about what the other has said. The fact that everyone will know this is possible, or at least could know it, doesn't negate the likelihood that false positives will happen at a much higher rate.

A good analogy here is the experience in health care when MRI scans first became available on a widespread basis.[13] A kind of natural experiment, it was a test of skilled peoples' ability to quickly incorporate awareness that there was no established baseline of what was 'normal' anatomic variation, or at least an observation not worth doing anything about, when suddenly the resolution at which one could see inside body structures jumped. In practice, false positives became common, as almost every MRI scan showed what looked like abnormalities against a now obsolete baseline (generally the much less detailed X-ray). What was later understood to be normal variation, was at the beginning over-interpreted as pathology – in other words, the signal was read as having more meaning than it actually contained. There were costly consequences – unnecessary procedures and surgeries that exposed patients to risks without corresponding benefits.

It's easy to foresee the same dynamic unfolding in the early years of widespread machine translation usage, and particularly in spoken language where peoples' concentration levels vary widely. I will think I understand what you are saying better than I really do, perhaps by taking your words too literally or filtering them through my cultural metaphors (or weakly understood versions of yours)). My false positives might sometimes be curiosities only, but they could mat-ter quite a lot for example if I create expectations about what you are going to do next based on my (faulty) understanding – and then you do things that defy those expectations in ways that hurt my interestsThe 'right' response to that scenario would be for me to re-correlate my confidence levels in the output of the translation algorithm from the start. The likely response (more likely in cases where you are an adversary or competitor) will be for me to infer that I've been deceived or manipulated by you, leading to a decline in ambient trust and higher levels of conflict.

---

[13] I lived through this as a medical student and it was really memorable and enlightening how many 'incidentalomas' we uncovered. Of course the inability to parse false positives wasn't just cognitive, it was also driven by codes of practice, insurance and potential malpractice liability concerns which took their own time catching up to the 'new' knowledge environment.

Over time, false positives will become less frequent as baseline understandings evolve in the presence of experience (exactly what happened with MRIs). But that process could easily take a decade or more, and it's the costs and instabilities associated with the transition period that will stand out.

***Diversity of Thought and the Long Tail***

In 2006, Chris Anderson's *The Long Tail* popularized the argument that the Internet's ability to reduce search costs and (for digital goods at least) the costs of holding nearly infinite inventory would lead to a flourishing of latent human tastes for an extremely broad diversity of products and ideas.[14] The iconic example was supposed be music and books. Instead of a small number of superstar titles dominating markets, the Internet would create a 'long tail' where small numbers of people had strong preferences for much greater diversity. The argument made intuitive sense and it slotted in nicely to idealistic visions of digital liberalism which were still common in the that decade.

Unfortunately it turned out to be mostly wrong.. Breaking down the long tail argument into its component parts helps to clarify important lessons that apply to machine translation.The fact that digital environments can support more variety than traditional physical markets, does not necessarily mean that they will do so. Consider the logic of the demand side. In theory, using Google instead of print or broadcast media to learn about rare products or ideas should enable much more personalized preference matching. But note the underlying assumption that there exist 'original' preferences which are in fact much more varied than physical search and distribution mechanisms were able to satisfy. Similarly in theory, sophisticated personalization algorithms that help people search (advanced versions of 'people who like X also like Y') should enhance the long tail effect. But note the underlying assumption here, that personalization algorithms are tuned to do that, rather than tuned to the contrary assumption that people like X and Y not because of some unique personal taste but because they want to consume ideas that can contribute to sociability with others who have consumed similar ideas.

---

[14] Anderson, C. (2006) *The long tail*. New York, NY: Hachette Books.

Brynjolfsson, E., Hu, Y. J., & Smith, M. D. (2010) The longer tail: The changing shape of Amazon's sales distribution curve. Retrieved from https://ssrn.com/abstract=1679991.

22 of 42

An alternative view is that baseline demand is actually *less* diverse than we might imagine, simply because most ideas are valued in a social context where people want to talk with each other. People might want to consume 'superstar content' simply because that act supports social interaction. They also might use popularity as a proxy for otherwise hard-to-measure quality.[15] Combine those two elements of the demand function and it's easy to see why superstars still dominate over the long tail and in many markets have actually increased their dominance.

The logic of the supply side might point in the same direction. What are the incentives to *create* a diversity of ideas? Technology certainly reduces the cost of 'stocking' and 'distributing' niche ideas, effectively to zero. But *creating* new ideas that matter is still hard and expensive. The expected return on investment in a niche idea remainsvery low unless and until it becomes popular.

Ben Thompson's aggregation theory explains how the current structure of major digital markets enhances this logic.[16] The decline in power of gate-keepers such as TV networks and traditional publishers who once controlled access to means for the promotion and distribution of ideas didn't create a level playing field that simply empowered individuals. Instead it gave to rise to aggregators who gained market power by pulling together and organizing demand, and then presenting that demand to suppliers. Search engines and social networks are the iconic examples.

---

[15] There is a significant individual cognitive burden associated with assessing new ideas; this limitation on the demand for diversity would operate independently of the desire for sociability around closely related ideas.

[16] Thompson, B. (2015, July 21). Aggregation Theory. Retrieved from https://stratechery.com/2015/aggregation-theory/

How would these dynamics change with machine translation? It's hard to imagine a mechanism by which reduction of language barriers within markets for ideas would re-establish power for supply-side gate keepers. Rather, it seems likely to reinforce the power of the demand-side aggregators, who now would be in a position to aggregate demand even more effectively from a much larger set of potential consumers regardless of their native language. To the extent that ideas seek large markets as products do, then the returns to a popular idea in a larger market with fewer language barriers will be even greater.[17] Demand-side aggregators will be the ones sending signals to suppliers about where those concentrations lie. Their power would be enhanced even more by the fact that some of the most important aggregators – including Google – are also the leaders in machine translation technologies. And so their investment and deployment strategies for machine translation could be tweaked in ways that enhance their power as aggregators even further.

The boldest hypothesis about how this would look at scale has three elements. First, the superstar phenomenon would converge into a small number of highly populated 'echo chambers' of ideas, reducing overall diversity and increasing the intensity of competition (and possibly conflict) among them. It would become harder for creative new coalitions to emerge, because it will be difficult for idea entrepreneurs to mobilize small groups with partially overlapping ideas and pull them together through compromise. Second, there would be enhanced incentives to create ideas that have the potential to become superstars – but vanishingly few would be able to do so and the vast majority would get little traction and likely die. Third, the strongest incentives will probably be to create 'complementary ideas' (as well as products) that naturally appeal to already existing superstar idea clusters (since these are the biggest pre-existing markets primed for complementary consumption).

---

[17] 'returns' could obviously be economic, or otherwise ('fame' and notoriety is not only and always a purely economic concept/value of course).

Another way to think about that argument is to highlight the attractiveness to idea entrepreneurs of marginally-related ideas and products that 'pile on' to existing echo chambers. From a risk-adjusted incentive standpoint, this would be a more rational approach for suppliers to take, than to try the higher-return but almost infinitely higher-risk alternative of creating a truly new idea. Overall, that doesn't bode well for innovation, pluralism, and other liberal values.

*Technical Barriers, Cultural Barriers, Differentiation*

Naive globalization arguments tend to ignore the extent to which societies and individuals harbour independent desires and demands for boundaries. Some of this comes from traditional motivations behind trade protectionism, and some is cultural, emotional and even religious.

When technology reduces or eliminates one barrier, other barriers rise in relative importance (sometimes even in absolute importance) as people intentionally find ways to 'protect' and differentiate themselves from unrestricted flows. When container shipping made it economically viable to concentrate the production of skis in the most efficient American factories, the Japanese claimed that snow in Japan was distinctive and required a different length of ski that was only made in Japan. These kinds of non-tariff barriers are common in culturally sensitive areas like food and the arts.[18] Non-tariff barriers are often criticized (sometimes ridiculed) as indirect ways to subvert trade liberalization but that is a value judgement which assumes economic efficiency as the primary goal of human interaction, which of course it often isn't. Minus any value assumptions, the question here is what kinds of new barriers will countries, firms and people erect to counter some of the boundary-breaking consequences of machine translation in the next decade? There's no way to fully anticipate the answer, but two kinds of barrier-raising seem almost inevitable, because they directly connect to basic organizational and individual motivations.

Consider first the role of technical standards and particularly data structures as an indication of how firms might re-create barriers to competition ('moats'.) Contemporary discussions about data portability notwithstanding, it is right now not straightforward to move a complex data set from one CRM architecture to another; or from one cloud service to a competitors'. When language barriers decline, the importance of these technical barriers rise in relative terms and might rise further through intentional action. Imagine then a world in which it is much easier to translate between Chinese and Russian, than it is to 'translate' between Salesforce and Oracle (really, between data stored in a Salesforce platform and an Oracle platform). That's barely a hypothetical given the challenges of interoperability at present; the interesting question is how much higher those boundaries rise. The same phenomenon should be expected for the Internet as a whole, where interoperability is already declining for related reasons.[19] These already-visible efforts are likely to be a taste of the creative experiments in barrier-raising that firms and governments develop over the next decade.

---

[18] Aggarwal, V. & Evenett, S. (2010). Financial crisis, 'new' industrial policy, and the bite of multilateral trade rules. *Asian Economic Policy Review*, 5(2), pp. 221-244

26 of 42

For individuals, the primordial desire to differentiate will probably yield even more creative strategies. One area where a significant demand for differentiation can be foreseen is in what the early 21st century 'global elite' will use to mark itself off and qualify its own members. It's interesting to note that the significance of language for elite differentiation is a modern phenomenon. The pre-bourgeois ruling classes of Europe were able to define themselves and cohere without a common language — when the King of England married a Spanish Princess, they didn't actually need to talk to each other very much. Noble marriage was a function of machiavellian politics and shared kinship, and so an illiterate nobility could still function as a nobility in pre-modern times. But that's not true for an industrial era bourgeoisie: their interactions and class consciousness depend in part on language, communication and the economic and cultural coordination that follows. Durkheim's concept of organic solidarity expresses the same mechanism through the division of labour.[20] As Benedict Anderson wryly put it, you can sleep with anyone, but you can only read some peoples' writing.[21] And if you want to coordinate a complex division of labour (and rule or at least control it enough to be an effective capitalist) it helps enormously to have a common or at least a closely translatable language.

---

[19] Early results from ongoing work by Nicolas Merrill to measure this phenomenon is reported at https://medium.com/cltc-bulletin/internet-fragmentation-beyond-free-and-closed-cb8b1dfcd16a

[20] Durkheim, E., & In Simpson, G. (1933). *Émile Durkheim on The division of labor in society*. New York: Macmillan.

[21] Anderson, B. (1983). *Imagined communities reflections on the origin and spread of nationalism* (pp. 76). New York, NY: Verso.

The ability to speak multiple languages fluently thus became a signal of elite differentiation across a broad swathe of the world. Anderson described it as a key point of connection between local colonial masters and the colonizing state, with bilingual elites in the colonies ruling monoglot populations and thus controlling the flow of information and authority from the metropole.[22] That function transformed during the post-colonial era into a meaningful signal of elite status (one among several of course). Poor and non-cosmopolitan commoners wouldn't generally have the economic need to learn multiple languages, nor the resources to study them. And they wouldn't get the value of cultural signaling that happens when you speak multiple languages in the first-class cabins of airplanes, which supports affiliative sorting in job and marriage markets.

When multiple languages were a luxury good that required significant investment of time and money to attain, the ability to speak and read them were signals that elites could use to identify each other. That signaling function has been in decline for a while (in part because of earlier technologies, which have made it easier and cheaper to learn languages). Machine translation will further reduce the signaling value toward a zero asymptote. Which – repeating the earlier point – will create a need and demand for other ways of signaling elite differentiation. There are plenty of possible paths for this demand to be expressed; precisely how that emerges is likely to be a bit of a surprise. It may very well turn out to be an even more exclusionary set of signals that are even harder for non-elites to attain, which (like some of the non-tariff barriers above) will tend to make the new borders *less* permeable than the old.

---

[22] Anderson, B. (1983). Official Nationalism and Imperialism. In *Imagined communities reflections on the origin and spread of nationalism* (pp. 83-111). New York, NY: Verso.

**Amplify the Upside**

Machine translation could make the global economy more fractured and unequal over the next decade. But if we get a few important things right, machine translation could instead contribute to significant gains in economic productivity; to new and broader political and cultural coalitions; to further ethnic and genetic hybridization of the human species; to beneficial and sustainable forms of immigration; and to more profound cultural understandings that break down dysfunctional barriers limiting human progress.

*Productivity*

Consider the possible productivity effects to start. The most straightforward mechanism is simply a next-generation globalization effect – as with container shipping in the 20th century, boundaries (in this case, linguistic boundaries) between markets would fall leading to higher levels of competition and greater potential scale for production and distribution. The classical argu-ment is that productivity gains emerge from the re-allocation of resources across sectors following opening to trade. Other mechanisms that boost productivity include enhancing competition which puts pressure on domestic producers to lower price margins, improve efficiency and make greater efforts at innovation; and increasing the quality and variety of intermediate inputs that are available to domestic producers.23

There is some evidence on the possible magnitude of this effect from previous studies that point to the importance of common language as a trade enabler. Studies of regional trade politics in the 1990s are one source of relevant models. Frankel et. al. (1995) developed a gravity model of trade to test for the effect of regional trade agreements.24 (Gravity models are built on the proposition that 'baseline' trade between 2 countries is proportionate to the product of their GDPs and inversely proportionate to the physical distance between them).25 They found an independent effect of regional trade groupings that are layered 'on top' of gravity model expectations, demonstrating that regional trade agreements (even though they often in the 1990s corresponded

---

23 Helpman, E., & Krugman, P. R. (1985). *Market structure and foreign trade: Increasing returns, imperfect competition, and the international economy*. Cambridge, MA: MIT Press.

Aghion, P., Bloom, N., Blundell, R., Griffith, R., & Howitt, P. (2005). Competition and innovation: An inverted-u relationship. *The Quarterly Journal of Economics*, *120(2)*, pp. 701–728

Grossman, G. & Helpman, E. (1991) Quality ladders in the theory of growth. *The Review of Economic Studies, 58,* pp.43–61

24 Frankel, J., Stein, E., & Wei, S. (1995). Trading blocks and the Americas: The natural, the unnatural, and the super-natural. *Journal of Development Economics, 47,* pp.61-95.  these were becoming an increasingly popular alternative to multilateral trade at the time

25 By baseline I mean that gravity models suggest a 'natural' economic determinant of trade, which is a useful baseline against which one can assess the impact of policy and other drivers.

closely with physical contiguity) did more than simply place a label on trade flows that would have been present in a world without policy. But they also found an intriguingly large causal effect associated with common language. Incorporating a dummy variable for 9 languages and comparing otherwise matched trade dyads that differ in language commonality, they found that two countries with strong linguistic ties tend to trade 65 per cent more than two countries that have similar gravity model characteristics but different primary languages. This is a large effect and there are possible confounding variables — for example, countries with common languages frequently had colonial relationships in the past that may explain part of the effect through other mechanisms. But even if colonial links explain half the variance, there's a significant effect for language commonality. And if machine translation were to quickly remove just one-half of the remaining language barrier effect between two markets, that would amount to roughly a 16 per cent expected increase in trade – still a massive impact.

A more recent study that makes use of a partial natural experiment broadly reinforces this estimate. Brynjolfsson et. al. (2018) assess what happened when EBay introduced limited machine translation on its trading platform in 2014.[26] Looking at the impact of English-Spanish translation on US exports to Spanish-speaking Latin American countries the authors found an increase of 17–21 per cent depending on the time windows in which the comparison is made. They found also that the increase is greater for differentiated products, for products with more words in their listing titles, for cheaper products and for less experienced buyers – which are consistent with the likely mechanism that machine translation is boosting trade by reducing search costs for buyers.[27]

---

[26] National Bureau of Economic Research. (2018, August). *Does machine translation affect international trade? evidence from a large digital platform* (NBER Working Paper 24917). Cambridge, MA: Brynjolfsson, E., Hui, X., & Liu, M.

One shouldn't make too much of the precise numbers that these models generate – they are rough estimates based on an imperfect baseline (gravity model) and an imperfect natural experiment (Ebay). But the fact that the estimates fall in a similar range – and a range that is larger but not wildly inconsistent with some previous estimates of the degree to which language barriers can be trade-inhibiting – is indicative of what machine translation will do to intensify trade, create larger more contestable markets and reduce matching frictions *in well resourced language dyads* at first.[28] The productivity boost could be significant, given recent IMF models estimating that a one per cent absolute decline in tariffs can increase total factor productivity by as much as two per cent.[29] Even if that estimate is high, keep in mind that a small increase in productivity which continues over a meaningful period of time creates substantially greater absolute wealth levels. Productivity is colloquially the gift that keeps on giving, but only to those that experience its improvement in this case through the reduction of language barriers where machine translation works well.

---

[27] Put differently, products that have higher search costs that would be more significant in a buying decision tend to benefit more from machine translation. This finding increases confidence that the mechanism of trade facilitation is as expected.

[28] See Lohmann, J. (2011). Do language barriers affect trade? *Economics Letters, 110*, pp.159–162. Lohmann derives a Language Barrier Index which is a measure of similarity between languages using categories from the World Atlas of Language. In his gravity model estimation, a 10 per cent increase in the Language Barrier Index (which of course represents a 10 per cent decrease in common linguistic features) is associated with a 7–10 per cent decrease in trade flows between two countries. On matching frictions and market efficiency see Goolsbee 2002 Wasn't sure which Goolsbee you were referring to.

[29] International Monetary Fund. (2016, March). *Reassessing the productivity gains from trade liberalization* (IMF Working Paper WP/16/77). Washington, DC: Ahn, J., Dabla-Norris, E., Duval, R. A., & Hu, B. A partially comparable study that assesses the (tricky) relationship between trade declines during the acute phase of the Global Financial Crisis and Great Recession around 2010 and declining productivity growth is Petersen Institute for International Economics. (2016, October). *Increased Trade: A Key to Improving Productivity* (Issue Brief 16-15). Hufbauer, G. C. & Ju, Z.

*Politics and Culture*

Coalition politics in domestic settings should be expected to shift as well, but with mixed consequences. At least two kinds of effects are foreseeable and though it's difficult to estimate magnitudes, they are still worth considering as logical mechanisms. The first would be a gradual but meaningful expansion of cross-linguistic and by implication cross-national political coalitions. From a 'borderless world' perspective where barriers to movement of money and goods across national boundaries have fallen dramatically in the last 50 years, it's remarkable in relative terms how few truly cross-national political movements have emerged alongside economic globalization. Consider for example the outbreak of populist movements in the second half of the 2010s or anti-capitalist movements (or at least anti-bank movements) in the first half of that decade. These were multi-domestic phenomenon more than transnational ones, appearing in a number of countries simultaneously and for parallel reasons, but never really joining together to create a transnational movement.

Of course language is not the only barrier here – political landscapes, concrete interests, cultural predilections, even electoral systems and rules also function as equivalents of non-tariff barriers for politics. But language is likely a meaningful part of what holds back cross-national coalition formation and certainly impedes the mechanism of learning from parallel movements in other countries. A thought experiment that points in the right direction is to imagine an alternative history in which Marxist movements across Europe in the 19th-20th centuries could communicate and interoperate without language friction. The parallel 2020s thought experiment might involve modern labour movements (particularly workers suffering the transnational shocks that will be associated with robotics), transnational climate coalitions and perhaps transnational religious movements as well.

A second effect would be to modify what Robert Putnam (1988) called two-level games, where a leader appeals to a particular constituency at home and quite a different constituency abroad using distinct and sometimes incompatible arguments.[30] A concrete example is Israeli Prime Minister Benjamin Netanyahu, whose speeches in Hebrew for his domestic Israeli audience often have a very different tone and message than his speeches in English which are aimed at an international audience and often at the American Jewish community. Observers of Netanyahu's two-level game strategy often marvel at why he doesn't suffer greater 'hypocrisy costs' from these inconsistencies – and the language barrier is certainly part of the explanation.[31] As in some of the prior arguments, politicians wouldn't simply give up on the ability to play multiple games at once and would try to adapt by developing new ways to segregate messages – though none would probably be as simple and effective as language barriers.

---

[30] Putnam RD. Diplomacy and Domestic Politics: The Logic of Two-Level Games. International Organization [Internet]. 1988;41 (3) :427-460.

[31] Hypocrisy costs is a loosely-derived game-theoretic term to describe the reputation hit a leader supposedly would suffer from visible demonstrations of hypocritical behaviour, making it harder for her to create credible commitments in the future.

A related and more macro-political dynamic would likely emerge around exclusionary communities that are primarily defined by ethnicity. Language barriers can be instrumentally useful in keeping appeals and arguments largely contained within ethnically-defined groups that speak the same language and that might not view those exclusionary arguments as racist, aggressive, or violence-inducing – while those outside the group probably would. As language boundaries fall and the range of constituents goes up, so does the range of opponents, enemies and disruptors who have immediate access to the message. It's hard to foresee which directional effect predominates, but if you start with the assumption that 'sunlight is a good disinfectant' for exclusionary appeals, machine translation is more likely to be a constraint on and net negative for harsh exclusionary arguments in ethnic politics.

The landscape for individuals choosing life-partners and creating families would shift as well. It seems intuitive that linguistic homogamy (marriage among two people who speak the same language) is a desirable characteristic in a spouse, but this does limit substantially the size of 'marriage markets' even if you assume that these markets are mainly local or at best national (that has become a less robust assumption over the last decade as internet dating has exploded in popularity).[32] The percentage of inter-ethnic marriages in the US increased to 10.2 per cent of households in 2016 from 7.4 per cent in 2012, showing just how quickly behaviors can change as perceived barriers decline.[33] And while language obviously is not the only remaining barrier in marriage markets (ethnicity, nationality, religion and culture intrude) a recent study from Switzerland (a multi-lingual country where there are both co-nationals and non-nationals, each with common or different languages) suggests that after spatial barriers, linguistic differences are the largest remaining obstacle to inter-ethnic and inter-national marriage.[34] A decline in linguistic homogamy would accelerate further the rise of inter-ethnic marriages, whose precise effects on the further normalization of multi-ethnic children and other economic and socio-cultural behaviours are almost certainly auspicious.[35]

---

[32] a 2019 paper estimates that 39 per cent of heterosexual and 60 per cent of same-sex American couples met online in 2017, up from two per cent in 1995 and 22 per cent in 2009. Rosenfeld, M, Thomas, R. J., & Hausen, S., (2019). Disintermediating your friends: How Online Dating in the United States displaces other ways of meeting. *Proceedings of the National Academy of Sciences* 116(36)

[33] data from US Census Bureau (https://www.census.gov/library/stories/2018/07/interracial-marriages.html) the 'barriers' that declined in this case are likely a combination of social norms, search costs (the same dynamic as internet dating noted above), but not legal barriers (at least since the 1967 Loving v. Virginia case ruled anti-miscegenation laws unconstitutional).

[34] followed by religious homogamy, age similarity and educational status.

Schroedter, J. & Roessel, J. (2016, July). *The Importance of Linguistic Homogamy in (Inter)Marriages: Insights from a Multilingual Country*. Paper Presented at the 3rd ISA Forum of Sociology, Vienna, Austria.

[35] See for example the literature survey by Furtado, D. & Trejo, S. (2012) Interethnic Marriages and their Economic Effects. In A. F. Constant & K. F. Zimmerman (Eds.), *International Handbook on the Economics of Migration* (pp.276–292). Northampton, MA: Edward Elgar Publishing

Another common life-pattern that would be impacted is migration – particularly 'voluntary' migrations that are less politically sensitive, such as during retirement. Somewhat like marriage markets, the geographic shape of demand for retirement migration is sensitive to language since retirees are less likely than young people to learn new languages even as they seek lower health care costs among other costs of living in other countries. Though the notion of 'retiring abroad' has garnered popular attention in the US over the last decade, it is still a rare decision, likely around two per cent. And the most popular countries for retirement migration by Americans are those where either English or (in the cases of Mexico and Japan) the original birth language of the retiree is spoken.[36] As with other complex social dynamics, machine translation by itself won't impact the cultural, inertial and other obstacles that restrain retirement migration but it would remove the language barrier – which could have meaningful short term effects as well as contribute to longer term reduction in more persistent barriers.

To try to make specific predictions about the magnitude of shifts in life patterns and cultural predilections would be reckless, since language is only one component of these behaviours and is interdependent with other causal drivers. Still, this survey of some logical effects demonstrates the scope of what language barrier-breaking technologies could do to amplify the upside of liberal progress. It's important to keep in mind that these effects would concentrate among high-resource language dyads, and impact at a lower level in less well-resourced dyads. Poorly-resourced dyads would see almost none of these effects in the short and medium term. They would then suffer in relative terms, as people and resources including goods and ideas are re-directed toward inter-operable language dyads (the equivalent of trade diversion effects on third parties that follow preferential tariff reductions).

---

[36] see for example Table 5.J11 of U.S. Social Security Administration, Annual Statistical Supplement, 2018, https://www.ssa.gov/policy/docs/statcomps/supplement/2018/5j.html#table5.j11, measuring United States citizens who receive social security abroad as about 1.5 per cent. That is an imperfect proxy – for example, a United States retiree living abroad might maintain a bank account inside the United States and have her social security check deposited there

That is a challenge that policy needs to address. As I argued in the introduction, there are distinct advantages to moving quickly and in anticipation of the deployment of next-generation machine translation systems. That argument should seem even more urgent after this survey of possible upside effects, since many of them would tend toward positive feedback loops in which progress yields accelerating progress, at least for a while. The longer we wait to re-structure incentives and practices around machine translation technology, the harder it will likely be to retro-fit for broad liberal objectives.

**What is to Be Done?**

Consider from a design perspective how machine translation technology might be developed, licensed, distributed and paid for. The leading platform firms at present have the largest incentives along with the greatest resources to make investments, but their incentives and business models will continue to skew toward high resource language pairs. In addition, intelligence and security agencies of large governments will surely invest in a few less well resourced languages of particular geopolitical interest to them. The irony is that medium and poorly resourced lan-guages look set to be left behind *unless* the people and countries that speak them are perceived as a security risk by US and China in particular. This isn't a good outcome overall for human welfare in the next decade, because it reinforces (and could exacerbate) existing inequalities both economic and cultural while undermining fairness and protection of the most vulnerable.

But technology needs to earn a return on investment. An obvious way to create greater balance among language dyads would be to subsidize the economic return on attention to low-resourced languages. Who might be motivated to fund such subsidies? Two possible models that suggest analogous rationales might be contemporary schemes that even out pharmaceutical investment and distribution, and 20th century schemes that funded rural phone service. Machine translation does have some universal human rights-type characteristics, along the lines of access to small-molecule drugs that have large fixed costs to discover but low variable costs to distribute and use. Even a minuscule surcharge on transactions made viable by machine translation in high-resource settings might be sufficient to fund such a scheme, without distorting the new value-creating possibilities that the technology will enable.

People who today argue that internet connectivity is a universal human right should consider whether language interoperability might be even more essential. Language interoperability will be a prerequisite to exercising in meaningful ways the internet connectivity right that they are promoting. This is where the rationale behind subsidies for universal basic phone service in the 20th century is a relevant analogy. On a global landscape even more so than a domestic one, the ability to join a network is worth very little if your language isn't functionally useful on the network. It might be a net negative if the only effective purpose of joining the network is to become a passive consumer of products (physical and cognitive) produced and distributed in other languages. It's a constructive stretch to imagine how existing international institutions like the Global Fund might extend their mission toward machine translation; or how the Global Fund model might be repurposed in a new institution combining governments and technology firms from the start.

Secondarily, it's important to consider *in advance* how to protect and compensate groups that come out as losers even in relative terms. The lessons of standard trade theory - and the failure of political elites to anticipate how those lessons would manifest in political economy during the last twenty years — are directly applicable. The populist backlash of the last decade didn't happen because political economy is silent on the challenge of relative losers from trade. The problem was that the presumed stabilization mechanism suggested by theory – using some of the surplus value created by opening trade to compensate losers and keep them on board politically – isn't natural or automatic, and wasn't enacted by political elites. Without digressing into blame stories about the sources of that failure, there's an obvious lesson to be had about setting up of institutionalized mechanisms to compensate losers early in the process of technology development and rollout. It's reasonable to assume (though hard to measure precisely) and even more reasonable to act as if the costs of compensation rise, the longer the delay. In my view, political and ethical recriminations about elite failures in the 2000s and 2010s to act in the context of that prior era's globalization technologies, would be better redirected toward efforts to anticipate and act on machine translation consequences right now. A constructive mindset would focus on not making exactly the same mistakes again.

Another reason to do this in advance would be to reduce at least one of the motivations for security threats to machine translation technologies. Adversarial machine learning attacks on translation systems aren't the stuff of science fiction, and the consequences of even small manipulations of the underlying algorithms could be devastating and hard to detect until considerable damage is done. An adversarial machine learning attack on translation would be a a new form of disinformation campaign, with a multiplier effect when combined with manipulated video and audio. Cybersecurity for machine translation systems is likely to become its own sub-speciality in the security world, and the luddite-type motivations of those who lose in relative terms is only one of many reasons why people might design attacks. But acknowledging that motivation and doing things to reduce it would also call attention to the needed upfront security investments aimed at hardening the underlying technologies and building awareness about threat and risk models in advance.

As in other cybersecurity and particularly adversarial machine learning issues, dealing with the technology itself will only go so far. Probably the most important and most challenging aspect for individuals and groups of people will be the risk of false positives in understanding, that I discussed in section 3 of this paper. Maintaining and extending diversity of thought and ideas in a world of widespread machine translation will be as important, though more abstract and harder to measure. The upsides of this technology revolution will be extraordinary and reveal themselves over time, but the transition period – which could be a decade or more – will have the most politically and socially salient and sensitive dilemmas to manage. Anticipation and pre-emptive management is by far the best strategy.